\documentclass[12pt]{article}
\usepackage{amsmath} 
\usepackage{amssymb}
\usepackage{bm}
\usepackage{bbm}
\usepackage{cite}
\usepackage{color} 
\usepackage{graphicx}
\usepackage{slashed} 
\usepackage{tabularx}
\usepackage{comment}

\newcommand{\Sec}[1]{Sec.~\ref{#1}}  

\newcommand{\fig}[1]{Fig.~\ref{#1}}

\newcommand{\Eq}[1]{Eq.~(\ref{#1})}
\newcommand{\eq}[1]{Eq.~(\ref{#1})}

\newcommand{\cmt}[1]{}
\begin{document}

\title{
\vspace{-1truecm}
\Large\bf
Sommerfeld enhancements with vector, scalar and pseudoscalar
force-carriers
}

\author{
 Ze-Peng Liu$^a$\footnote{Email: zpliu@itp.ac.c}, \ 
 Yue-Liang Wu$^{a,b}$\footnote{Email: ylwu@itp.ac.cn}, \ 
 and  Yu-Feng Zhou$^a$\footnote{Email: yfzhou@itp.ac.cn}
  \\  \\
  \textit{$^a$State Key Laboratory of Theoretical Physics},\\
\textit{Kavli Institute for Theoretical Physics China}, \\
  \textit{Institute of  Theoretical Physics, Chinese Academy of Sciences}\\
  \textit{Beijing, 100190, P.R. China}\\
   \textit{$^b$University of Chinese Academy of Sciences},\\
    \textit{Beijing, 100049, P.R. China}\\
}
\date{}
\maketitle
\begin{abstract}
The first AMS-02 measurement confirms the existence of 
an excess in the cosmic-ray positron fraction previously 
reported by the PAMELA and Fermi-LAT experiments.
If interpreted in terms of  thermal dark matter (DM) annihilation,
the AMS-02 result  still suggests that 
the  DM annihilation cross section in the present day should be
significantly larger than that at  freeze out. 
The Sommerfeld enhancement of 
DM annihilation cross section is 
a possible explanation, which is however
subject to the  constraints from  DM thermal relic density,
mainly due to the annihilation of  DM particles into  force-carrier particles 
introduced by the mechanism. 
We show that 
the effects of the Sommerfeld enhancement and 
the relic density constraints depend  significantly on 
the nature of the force-carrier.
Three scenarios where the force-carrier
is a vector boson, scalar and pseudoscalar particle are investigated
and compared.
The results show that
for the case with vector  force-carrier, 
the Sommerfeld enhancement
can marginally account for the AMS-02 data
for DM particle annihilating into $2\mu$ final states, 
while for scalar force-carrier the allowed Sommerfeld enhancement factor
can be  larger by a factor of two. 
For the case with a pesudoscalar force-carrier, 
the Sommerfeld enhancement factor can be
very large in the resonance region, 
and it is possible to  accommodate  the AMS-02 and Fermi-LAT 
result for a variety of DM annihilation final states.
\end{abstract}
\newpage

\section{Introduction}
Evidence from 
astronomical observations at different scales has indicated that
dark matter (DM) contributes to nearly 26\% of the energy density  of 
the Universe~\cite{Ade:2013xsa,
Ade:2013zuv 
}.
Popular DM candidates such as 
the weakly interacting massive particles (WIMPs) are 
expected to annihilate or decay into 
standard model (SM) final states in the Galactic halo and beyond, 
which may leave imprints in the fluxes of cosmic-ray particles.

In the recent years, 
the PAMELA  collaboration has reported  a sharp upturn  of
the ratio of the positron flux to the total flux of electrons and positrons 
 in the energy range $\sim10-100$ GeV,
which is in excess over a  conventional astrophysical background 
\cite{Adriani:2008zr, 
1001.3522 
}, 
and was  confirmed by the Fermi-LAT data up to $\sim 200$ GeV
\cite{1109.0521 
}.
The total fluxes of electrons and positrons  measured by 
ATIC \cite{Chang:2008zzr 
}
and  Fermi-LAT~\cite{Abdo:2009zk,
Ackermann:2010ij 
}  
also showed possible excesses over
the expectations of  the conventional  background.
Recently, 
the AMS-02 collaboration released  the first measurement  of 
the positron fraction  
with unprecedented accuracy
\cite{PhysRevLett.110.141102
}.
Although the AMS-02 data are  consistent with   
the previous measurements of PAMELA, 
the measured spectrum from AMS-02  is slightly lower than that from PAMELA
for electron/positron energy higher than $\sim40$ GeV,
and  the slope of the positron fraction 
decreases  by an order of magnitude from $\sim20$ GeV to $\sim250$ GeV.
The implications of the precision AMS-02 data on the DM annihilation 
have been discussed (see e.g. Refs.~\cite{
Kopp:2013eka,
DeSimone:2013fia,Yuan:2013eja,
Cholis:2013psa,Jin:2013nta,
Yuan:2013eba,Kajiyama:2013dba,
Feng:2013vva
}). 
Several fits to the AMS-02 data showed that if $2\mu$ is the dominant DM annihilation channel,  
the AMS-02 favoured DM particle mass is $\sim 400-500$ GeV, 
and the thermally averaged product of annihilation cross section and relative velocity is 
$\langle\sigma v_{\text{rel}}\rangle \sim 10^{-24}\text{ cm}^{3}\text{ s}^{-1}$
\cite{DeSimone:2013fia,Feng:2013vva,Jin:2013nta}.
For instance, in Ref.~\cite{Jin:2013nta}, 
using a conventional astrophysical background,
the best fitted DM particle mass is found to be  $m_{\chi}\approx 460$ GeV,
with an annihilation cross section 
$\langle\sigma v_{\text{rel}}\rangle \approx 1.9\times 10^{-24}\text{ cm}^3\text{s}^{-1}$.
The DM annihilating into $2e$ is not favoured as the predicted positron spectrum is too hard.
For $2\tau$ final states, 
the favoured DM particle mass is $\sim 1.4$ TeV and 
the annihilation cross section $\sim 1.7 \times 10^{-23}\text{ cm}^3\text{s}^{-1}$, which is
compatible with the Fermi-LAT data \cite{Jin:2013nta}. 
Note that the $2\tau$ final states can generate large flux of diffuse gamma rays which 
is stringently constrained by the current observations.
Although  it seems that 
the favoured parameter regions by the current AMS-02 data are
different from that by Fermi-LAT for some leptonic final states,
all the current experimental data suggest that
the DM annihilation cross section in the present day  must be  
larger than  the typical WIMP thermal cross section 
$\langle \sigma v_{\text{rel}}\rangle_{F}\approx 3 \times 10^{-26} \text{ cm}^{3} \text{s}^{-1}$ at freeze out, 
which calls for nonstandard nature of DM particles.

The Sommerfeld enhancement has been considered as 
a mechanism which can naturally enhance the DM annihilation cross section 
at low relative velocities~\cite{
Sommerfeld31,Hisano:2002fk,Hisano:2003ec,Cirelli:2007xd,ArkaniHamed:2008qn,
Pospelov:2008jd,MarchRussell:2008tu,Iengo:2009ni,Cassel:2009wt}.
(for other mechanisms, see e.g. \cite{Feldman:2008xs,Ibe:2008ye,Guo:2009aj,Liu:2011aa,Liu:2011mn}).
In this scenario, 
the cross section of the DM annihilation 
$\bar\chi\chi\to X \ (X=2\mu, 4\mu, \dots)$ is velocity-dependent,
due to the multiple exchange of some light force-carrier particle $\phi$ between 
the annihilating DM particles $\bar\chi\chi$.
The thermally averaged annihilation cross section 
can be close to  $\langle\sigma v_{\text{rel}}\rangle_F$  
at the time of  thermal freeze out,  
but becomes much larger now  as 
the temperature of the Universe  in the present day 
is much lower.
Constraints on the Sommerfeld enhancement can be  obtained from
astrophysical observations 
(see e.g. Refs.
\cite{
Zavala:2009mi,Hannestad:2010zt,
Hisano:2011dc,Kamionkowski:2008gj,
Bovy:2009zs,Buckley:2009in,Feng:2009hw,
Cholis:2010px,Lattanzi:2008qa,
Robertson:2009bh,Cirelli:2010nh,Abazajian:2011ak
}). 
Among them, a stringent  constraint on the Sommerfeld enhancement
can arise from the  DM thermal relic density itself, 
which is less sensitive to the astrophysical uncertainties.
This is due to the fact that,  
in this mechanism,
the DM particles  inevitably annihilate into the force-carriers through 
the process like $\bar\chi\chi\to \phi\phi$, 
which enhances the DM total annihilation cross section at freeze out, 
and the relevant parameters are constrained by the DM relic density.
This reduces the allowed values of the Sommerfeld enhancement factor at lower temperature
\cite{Feng:2009hw,Feng:2010zp}.
Based on a  model in which $\phi$ is a $U(1)$ vector gauge boson,
it has been illustrated  that 
under  the relic density constraint,
the Sommerfeld enhancement factor is not large enough to account for 
the excesses reported by the PAMELA and Fermi-LAT experiments
\cite{Feng:2009hw,Feng:2010zp}.   
Note that the Sommerfeld enhancement can be realized with 
different type of force-carriers, 
such as scalar and pseudoscalar particles
\cite{ArkaniHamed:2008qn,MarchRussell:2008yu,Bedaque:2009ri}.
An $U(1)$ vector gauge bosons can be naturally light under the protection 
of gauge symmetry.  A pseudoscalar particles can also be naturally light if
they play the role of  a pseudo-Goldstone boson. A light scalar particle can
be stable with the help of  supersymmetry.  
The effect of the Sommerfeld enhancement and 
the constraint from thermal relic density depend on
the nature of the force-carrier particle. 
For instance, if $\phi$ is a scalar particle, 
the cross section for  $\bar\chi\chi\to \phi\phi$ is velocity suppressed,
resulting in a weaker constraint 
compared with  the case where $\phi$ is a vector boson.
If $\phi$ is a pseudoscalar, 
the induced potential is of tensor force type rather than the Yukawa type.

In light of the recent AMS-02 results, 
it is of interest to investigate whether  
the Sommerfeld enhancement can account for  
the more accurate AMS-02 data 
in generic cases.  
In this work,
we explore  and compare the  Sommerfeld enhancements 
with three different type of force-carriers:  vector, scalar and pesudoscalar,
under the constraint from DM thermal relic density.
We show that 
for vector boson force-carrier 
the Sommerfeld enhancement  can only 
marginally account for the AMS-02 data, 
for scalar force-carrier 
the  allowed Sommerfeld enhancement factor can be larger roughly by a factor of two,
while in the case of pesudoscalar force carrier, 
much larger enhancement can be obtained in the resonance region. 

This paper is organized as follows. 
In \Sec{sec:smf}, 
we outline the  formalism  of the Sommerfeld enhancement and 
the thermal evolution of the DM number density.
In \Sec{sec:constraint}, 
we discuss the  Sommerfeld enhancement and the constraints 
for the cases with vector, scalar and pseudoscalar force carriers,
and compare the allowed enhancement factors with 
the current experimental data.  
The nature of Sommerfeld enhancement  with pseudoscalar is discussed in detail.
The conclusions are given in \Sec{sec:conclusion}.

\section{Mechanism of Sommerfeld enhancement}\label{sec:smf}
\begin{figure}[thb]
\begin{center}
\includegraphics[width=0.45\textwidth]{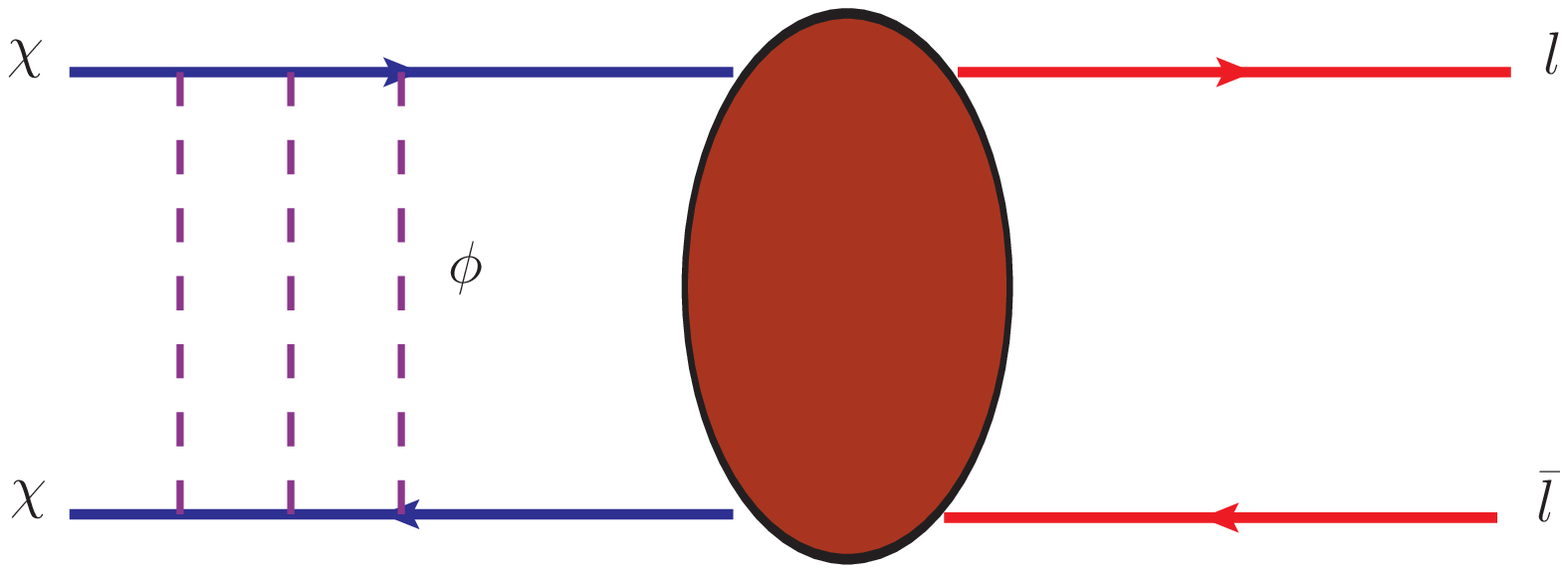}
\includegraphics[width=0.4\textwidth]{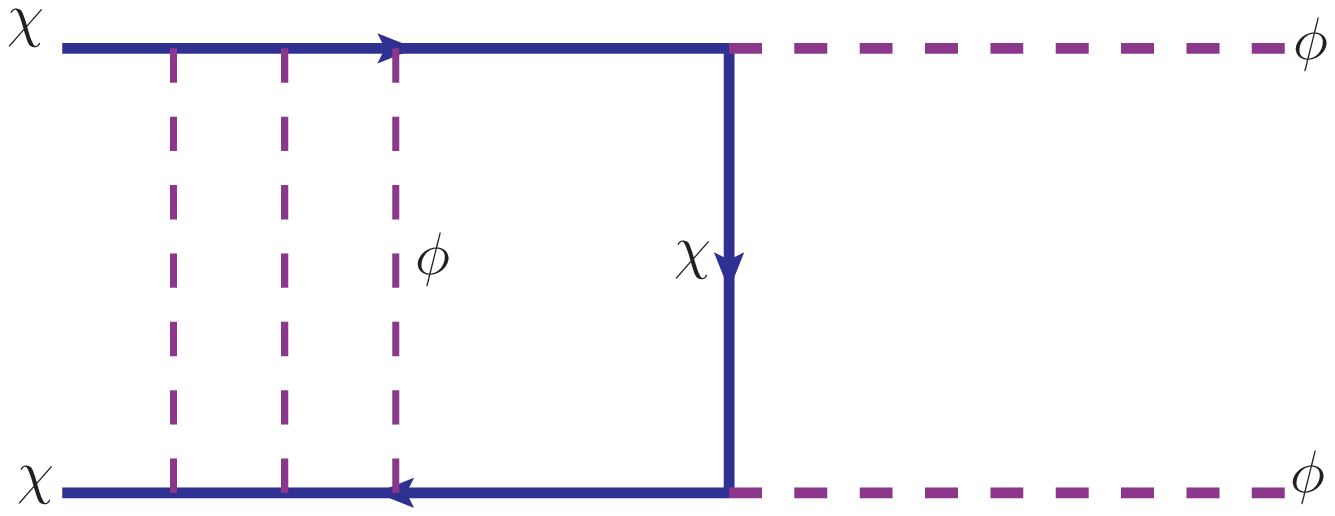}
\end{center}
\caption{
(Left) Feynman diagram of DM annihilation process $\bar\chi\chi\to X$, 
$(X=2\mu, 2\tau,\dots)$ with multiple force-carrier exchange which results in Sommerfeld enhancement
of the annihilation cross section.
(Right) Feynman diagram of DM annihilation into the force-carriers through 
$t$-channel process $\bar{\chi}\chi\rightarrow 2\phi$.}
\label{fig:diagrams}
\end{figure}

The Sommerfeld enhancement of DM particle annihilation cross section occurs 
when the annihilating particles  self-interact through a 
long-range attractive potential  $V(\mathbf{r})$ at low relative velocities \cite{Sommerfeld31}. 
In this scenario, 
the short-distance DM particle annihilation cross section can be greatly enhanced due to
the distortion of the wave function of the annihilating particles at the origin
\cite{Hisano:2002fk,Hisano:2003ec,Hisano:2004ds,Cirelli:2007xd}. 
The attractive potential can be induced from 
the multiple-exchange of some light force-carrier particle $\phi$ between 
the annihilating DM particles as shown in the left panel of \fig{fig:diagrams}. 
The nature of Sommerfeld enhancement has been
extensively studied (see  e.g. 
Refs.~\cite{ArkaniHamed:2008qn,Iengo:2009xf,Iengo:2009ni,Cassel:2009wt,Slatyer:2009vg,Hryczuk:2010zi,
Dent:2009bv,Zavala:2009mi,Feng:2009hw,Feng:2010zp,
Finkbeiner:2010sm,Hryczuk:2011vi,McDonald:2012nc})
in light of the cosmic-ray positron/electron excesses reported  by
PAMELA~\cite{Adriani:2008zr}, ATIC \cite{Chang:2008zzr}, and Fermi-LAT
\cite{Abdo:2009zk} etc..

The effect of Sommerfeld enhancement can be described by the following 
non-relativistic Schr$\ddot{\mbox{o}}$dinger equation for 
the  two-body wave function $\Psi(\mathbf{r})$ of the annihilating DM particles
\begin{equation}
  -\frac{1}{m_{\chi}}\nabla^{2}\Psi(\mathbf{r})+V\left(\mathbf{r}\right)\Psi\left(\mathbf{r}\right)
  =
  \frac{m_{\chi}v_{\text{rel}}^{2}}{4}
  \Psi\left(\mathbf{r}\right),
\end{equation}
where $\mathbf{r}$ and  $v_{\text{rel}}$ are the relative distance and velocity of 
the two annihilating  DM particles, respectively.
After an expansion over the Legendre polynomial $P_{\ell}(\cos\theta)$ with
angular momentum $\ell$, namely,
$\Psi(r,\theta)=\sum_{\ell}P_{\ell}(\cos\theta)\chi_{\ell}(r)/r$, 
with $r=|\mathbf{r}|$ and $\theta$ the zenith angle of spherical coordinates, 
the Schr$\ddot{\mbox{o}}$dinger equation for radial wave function $\chi_{\ell}(r)$
can be written as
\begin{equation}\label{eq:shrodinger}
\frac{d^{2}\chi_{\ell}\left(r\right)}{dr^{2}}
-\sum_{\ell'}
\left[
m_{\chi} V_{\ell\ell'}(r)
+\frac{\ell\left(\ell+1\right)}{r^{2}}\delta_{\ell\ell'} 
\right]\chi_{\ell'}\left(r\right)
+k^{2} \chi_{\ell}\left(r\right)=0,
\end{equation}
where $k\equiv m_{\chi} v_{\text{rel}}/2$ and $V_{\ell \ell'}$ is given by 
\begin{eqnarray}\label{eq:pw_seq}
V_{\ell\ell'}\left( r \right) 
& = & 
\frac{\left(2\ell+1\right)}{2}
\int_{-1}^{+1}P_{\ell}\left(\cos\theta\right)V\left(r,\theta\right)P_{\ell'}\left(\cos\theta\right)    d(\cos\theta) .
\end{eqnarray}
The above Schr$\ddot{\mbox{o}}$dinger equation can be
solved with the following boundary conditions \cite{Iengo:2009ni,Cassel:2009wt}
\begin{eqnarray}\label{eq:boundary}
\lim_{r\rightarrow0}\chi_{\ell}\left(r\right)  =  (k r)^{\ell+1} \ \text{and}\
\lim_{r\rightarrow0}\frac{d\chi_{\ell}\left(r\right)}{dr}  =  k(\ell+1)(k r)^{\ell}  .
\end{eqnarray}
The asymptotic behaviour of the wave function at infinity is
\begin{equation}\label{eq:asymptotic}
\lim_{r\rightarrow\infty}\chi_{\ell}\left(r\right)  
\rightarrow 
C_{\ell}\sin\left(kr-\frac{\pi}{2}\ell+\delta_{\ell}\right),
\end{equation}
where $\delta_{\ell}$ is the phase shift and $C_{\ell}$ is a normalization constant. 
With the aforementioned  boundary conditions, 
the Sommerfeld enhancement factor $S_{\ell}$  for 
a partial wave $\ell$ is given by
\cite{ArkaniHamed:2008qn,Iengo:2009ni}
\begin{equation}
 S_{\ell}\equiv
 \lim_{r\to0}\left|\frac{\chi_{\ell}\left(r\right)}{\chi_{\ell}^{(0)}\left(r\right)}\right|^{2}
=\left[\frac{(2\ell+1)!!}{C_{\ell}} \right]^{2} ,
\end{equation}
where $\chi_{\ell}^{(0)}(r)$ is the wave function in the free-motion case without a potential.

The exchange of a massive vector or scalar particle $\phi$ with mass $m_{\phi}$ 
between the DM particles results in an attractive Yukawa potential
\begin{equation}\label{eq:Yukawa}
  V(r)=-\frac{\alpha e^{-m_{\phi}r}}{r},
\end{equation}
where $\alpha$ is the coupling strength.
In the limit of $4\alpha m_{\phi}/m_{\chi}\ll v_{\text{rel}}^{2}$, 
the Yukawa potential in the Schr$\ddot{\mbox{o}}$dinger equation
can be well approximated by a Coulomb-type potential,
and the Schr$\ddot{\mbox{o}}$dinger equation can be solved analytically for
arbitrary angular momentum.
The enhancement factors read~\cite{Cassel:2009wt}
\begin{equation}\label{eq:smf_com}
S_{0}(v_{\text{rel}}) \approx  
\left(\frac{2\pi\alpha}{v_{\text{rel}}}\right)
\frac{1}{1-e^{-2\pi\alpha/v_{\text{rel}}}}, \quad
\text{and} \quad
S_{1}(v_{\text{rel}}) \approx S_{0}(v_{\text{rel}})
\left(1+\frac{\pi^{2}\alpha^{2}}{{v_{\text{rel}}}^{2}}\right).
\end{equation}
Therefore, at low velocities, the $s$- and $p$-wave Sommerfeld enhancement factors scale as
$v_{\text{rel}}^{-1}$ and $v_{\text{rel}}^{-3}$ respectively.
In the case where $m_{\phi}$ is non-negligible, the $v^{-1}_{\text{rel}}$ behavior of
 $s$-wave cross section breaks down. Through approximating the Yukawa potential by the
Hulth$\acute{\mbox{e}}$n potential, the $s$-wave Sommerfeld enhancement factor
can be estimated as \cite{Cassel:2009wt,Slatyer:2009vg}
\begin{equation}\label{eq:smf_ykw}
S_{0}(v_{\text{rel}}) \approx \left(\frac{2\pi\alpha}{v_{\text{rel}}}\right)\frac{\sinh\left(\frac{6v_{\text{rel}}m_{\chi}}{\pi m_{\phi}}\right)}{\cosh\left(\frac{6v_{\text{rel}}m_{\chi}}{\pi m_{\phi}}\right)-\cos\left(\sqrt{\frac{24\alpha m_{\chi}}{m_{\phi}}-\frac{36m_{\chi}^{2}v_{\text{rel}}^{2}}{\pi^{2}m_{\phi}^{2}}}\right)}. 
\end{equation}
For $4\alpha m_{\phi} \gg v_{\text{rel}}^{2}$,
namely, 
the deBroglie wavelength of incoming particles is much
longer than the range of the interaction, 
the $s$-wave Sommerfeld enhancement saturates with $S_{0}\sim 12/\epsilon_{\phi}$
where $\epsilon_{\phi}\equiv m_{\phi}/(\alpha m_{\chi})$.  
But for some particular values of $\epsilon_{\phi}\simeq 6/(\pi^{2}n^{2}), (n=1,2,3,\dots)$ at  which 
the DM particle can form zero-energy bound states, 
there exists additional resonant enhancements which scale as $v_{\text{rel}}^{-2}$. The resonant
enhancement is eventually cut off by the finite width of the resonance
\cite{ArkaniHamed:2008qn}.

The velocity dependence of $p$-wave enhancement was investigated in 
Refs. \cite{Iengo:2009xf,Iengo:2009ni,Cassel:2009wt,Tulin:2013teo}.
Its  effect on the freeze out and thermal relic density was studied in 
detail in Ref.~\cite{Chen:2013bi}.
The generic $p$-wave annihilation
cross section before including the Sommerfeld enhancement is proportional to
$v_{\text{rel}}^{2}$.
Thus the velocity dependence of the total Sommerfeld-enhanced
$p$-wave annihilation cross section should be proportional to
$S_{1} {v^{2}_{\text{rel}}}$.  
As shown in Ref.~\cite{Chen:2013bi},
in the region where $\epsilon_{\phi} \lesssim 10^{-3}$, 
the total annihilation cross section 
scales as $v_{\text{rel}}^{-1}$ instead of $v_{\text{rel}}^{-3}$.  
In the resonance region $10^{-3}\lesssim\epsilon_{\phi} \lesssim 10^{-1}$, 
the velocity dependence of $S_{1}v_{\text{rel}}^{2}$ is not significant. 
In the saturation region $\epsilon_{\phi} \gtrsim 10^{-1}$, $S_{1}v_{\text{rel}}^{2}$ scales as $v_{\text{rel}}^{2}$,
the total cross section decreases rapidly towards low velocities. 
Thus the main difference from the $s$-wave case is that the total $p$-wave annihilation
cross section can be either velocity-suppressed or velocity-enhanced, depending
on the values  of $\epsilon_{\phi}$. 

The generic DM annihilation
cross section times the relative velocity before including the Sommerfeld
enhancement has the form $(\sigma
v_{\text{rel}})_{0}=a+bv_{\text{rel}}^{2}+\mathcal{O}(v_{\text{rel}}^{4})$,
where $a$ and $b$ are coefficients corresponding to the $s$- and $p$-wave
contributions which are assumed to be velocity-independent. After including the Sommerfeld
enhancement,
the thermally averaged cross section at a temperature $T$  or  $x \equiv m_{\chi}/T$ can be written as
\begin{equation}
\left\langle \sigma v_{\text{rel}}\right\rangle(x) 
=
a\langle S_{0}(v_{\text{rel}})\rangle(x)+b\langle S_{1}(v_{\text{rel}})v_{\text{rel}}^{2}\rangle(x),
\end{equation}
where the thermal average of a quantity $\mathcal{X}(v_{\text{rel}})$ in the non-relativistic limit is given by
\begin{equation}
  \left\langle \mathcal{X}\right\rangle(x) =\frac{x^{3/2}}{2\sqrt{\pi}}\int_{0}^{\infty} \mathcal{X}(v_{\text{rel}})
  e^{-\frac{xv_{\text{rel}}^{2}}{4}}v_{\text{rel}}^{2}\ dv_{\text{rel}}  .
\end{equation}
Due to the Sommerfeld enhancement, 
the thermally averaged annihilation cross section 
$\left\langle \sigma v_{\text{rel}}\right\rangle(x) $
depends on the parameters $\alpha$ and $m_{\phi}$.

The temporal  evolution of the DM number density  is governed
by the Boltzmann equation
\begin{equation} \label{eq:Boltzmann2}
\frac{dY}{dx}=-\sqrt{\frac{\pi}{45}}m_{\text{Pl}}m_{\chi}\frac{g_{*s} g_{*}^{-1/2}}{x^{2}}\left\langle \sigma v_{\text{rel}}\right\rangle \left[ Y^{2}-(Y^{\text{eq}})^{2}\right] ,
\end{equation}
where $Y^{(\text{eq})}\equiv n_{\chi}^{(\text{eq})}/s$ is the (equilibrium)
number density rescaled by entropy density $s$,
$m_{\text{Pl}}\simeq1.22\times10^{19}\text{ GeV}$ is the Planck mass scale.
$g_{*s}$ and $g_{*}$ are the effective relativistic degrees of freedom for
entropy and energy density, respectively.

The DM number density in the present day
can be obtained by integrating  Eq.~(\ref{eq:Boltzmann2}) with respect to $x$ in the region
$x_{f}\leq x \leq x_{\text{now}}$,  where $x_{f}\approx 25$ is the decoupling temperature, 
and $x_{\text{now}}\approx 4\times 10^{6}$ corresponds to the  temperature of halo DM in 
the present day
\begin{eqnarray}
  \frac{1}{Y\left(x_{\text{now}}\right)}
  & = & \frac{1}{Y\left(x_{f}\right)}+\sqrt{\frac{\pi}{45}}\ m_{\text{Pl}}\ m_{\chi}\int_{x_{f}}^{x_{s}}\frac{g_{*s}g_{*}^{-1/2}}{x^{2}}
  \left\langle \sigma v_{\text{rel}}\right\rangle dx .
\end{eqnarray}
Finally, the relic abundance of DM particles is given by
$ \Omega h^{2}\approx2.76\times10^{8} Y \left(x_{\text{now}}\right)
(m_{\chi}/\text{GeV})$, 
 which is to be compared with the observed value~\cite{Ade:2013zuv}
 \begin{equation}\label{eq:omegahExp}
( \Omega h^{2} )_{\text{exp}}=0.1187\pm 0.0017 .
 \end{equation}
In this work, we solve the \eq{eq:Boltzmann2}  directly using numerical approaches.
\section{Sommerfeld enhancement and relic density constraints
with different force-carriers}\label{sec:constraint}

The presence of Sommerfeld enhancement modifies 
the calculation of the DM thermal relic density. 
First, 
the Sommerfeld-enhanced cross section of  $\bar\chi\chi\to X$ increases 
with $x$ during freeze out,
which postpones the  decoupling of DM particles from the thermal bath and 
results in a decrease of the DM relic density~\cite{Dent:2009bv,Zavala:2009mi}.
Second, 
as the force-carrier is  much lighter than the DM particle, 
i.e., $m_{\phi}\ll m_{\chi}$, 
the DM particles necessarily annihilate into the force-carriers.
The process like $\bar\chi\chi\to \phi\phi$ will contribute to 
an annihilation channel in addition to $\bar\chi\chi\to X$, 
and can even be the dominant contribution to 
the total DM annihilation cross section,
which further reduces the DM relic density.
Thus in order to reproduce the observed DM relic density,
the relevant parameters such as the coupling $\alpha$ has to 
be small enough, which results in a reduction of the 
Sommerfeld enhancement factors at low temperatures~\cite{Feng:2009hw,Feng:2010zp}.
Before switching on the effect of Sommerfeld enhancement,
the total DM annihilation cross section $ (\sigma_{\text{tot}}v_{\text{rel}})_0$ 
can be written as the sum of the two contributions, namely,
$(\sigma_{\text{tot}} v_{\text{\text{rel}}})_{0} =(\sigma_{X}v_{\text{rel}})_{0}+(\sigma_{\phi\phi}v_{\text{rel}})_{0}$.
The thermally averaged total annihilation cross section after 
including the Sommerfeld enhancement has the form
\begin{align}\label{eq:sigmaTotal}
\langle \sigma_{\text{tot}} v_{\text{rel}}\rangle(x)
=
\langle S_{0}(v_{\text{rel}}) \rangle(x)  (\sigma_{X} v_{\text{rel}})_{0}
+
\langle S(v_{\text{rel}})(\sigma_{\phi\phi} v_{\text{rel}})_{0} \rangle(x), 
\end{align}
where $S(v_{\text{rel}})=S_{0(1)}(v_{\text{rel}})$, if the annihilation  $\bar\chi\chi\to \phi\phi$  proceeds through
$s(p)$-wave. 
In order to achieve the maximal Sommerfeld enhancement factor,
we have assumed that $\bar\chi\chi\to X$ is an $s$-wave process, 
and both $X$ and the decay products of $\phi$ are dominated by
SM charged leptons.
The boost factor of the DM annihilation is defined as 
\begin{align}\label{eq:boostFac}
B \equiv
\left( \frac{\rho}{\rho_{0}}\right)^{2}
\frac{\langle \sigma_{\text{tot}} v_{\text{rel}}\rangle(x_{\text{now}})}{\langle \sigma v_{\text{rel}}\rangle_{F}}  ,
\end{align}
where $\rho$ is the DM local energy density, and 
$\rho_{0}\approx 0.4 \text{ GeV}\text{cm}^{-3}$ is the 
DM energy density estimated from smooth DM density
profiles. 
In this work, we do not consider the boost factor from the local
clumps of substructure, namely, $\rho\approx \rho_{0}$ is 
assumed.
We parametrize the unknown cross section 
$\langle S_{0}(v_{\text{rel}}) \rangle(x)  (\sigma_{X} v_{\text{rel}})_{0}$ at freeze out
as 
$\langle S_{0}(v_{\text{rel}}) \rangle(x_{f})  (\sigma_{X} v_{\text{rel}})_{0}
\equiv
\eta \langle \sigma v_{\text{rel}}\rangle_{F}$. 
The boost factor can be rewritten as
\begin{align}
B\approx \eta S_{\text{eff}} +\frac{\langle S(v_{\text{rel}})(\sigma_{\phi\phi} v_{\text{rel}})_{0} \rangle(x_{\text{now}})}{\langle \sigma v_{\text{rel}}\rangle_{F}},
\end{align} 
where $S_{\text{eff}} \equiv \langle S_{0}(v_{\text{rel}}) \rangle(x_{\text{now}})/\langle S_{0}(v_{\text{rel}}) \rangle(x_{f})$ is 
the present-day $s$-wave Sommerfeld enhancement relative to that at freeze out.
The Sommerfeld enhancement factors $\langle S_{0,1}\rangle$ 
and the cross section $(\sigma_{\phi\phi} v_{\text{rel}})_{0} $
depend on 
the parameters $\alpha$ and $\eta$.
The requirement of  reproducing the correct thermal relic density
constrains  the  sizes  of $\alpha$ and $\eta$,
which will in turn limit the maximally allowed boost factor $B$.

\subsection{Vector boson force-carrier}
\begin{figure}[thb]
\begin{center}
\includegraphics[width=0.48\textwidth]{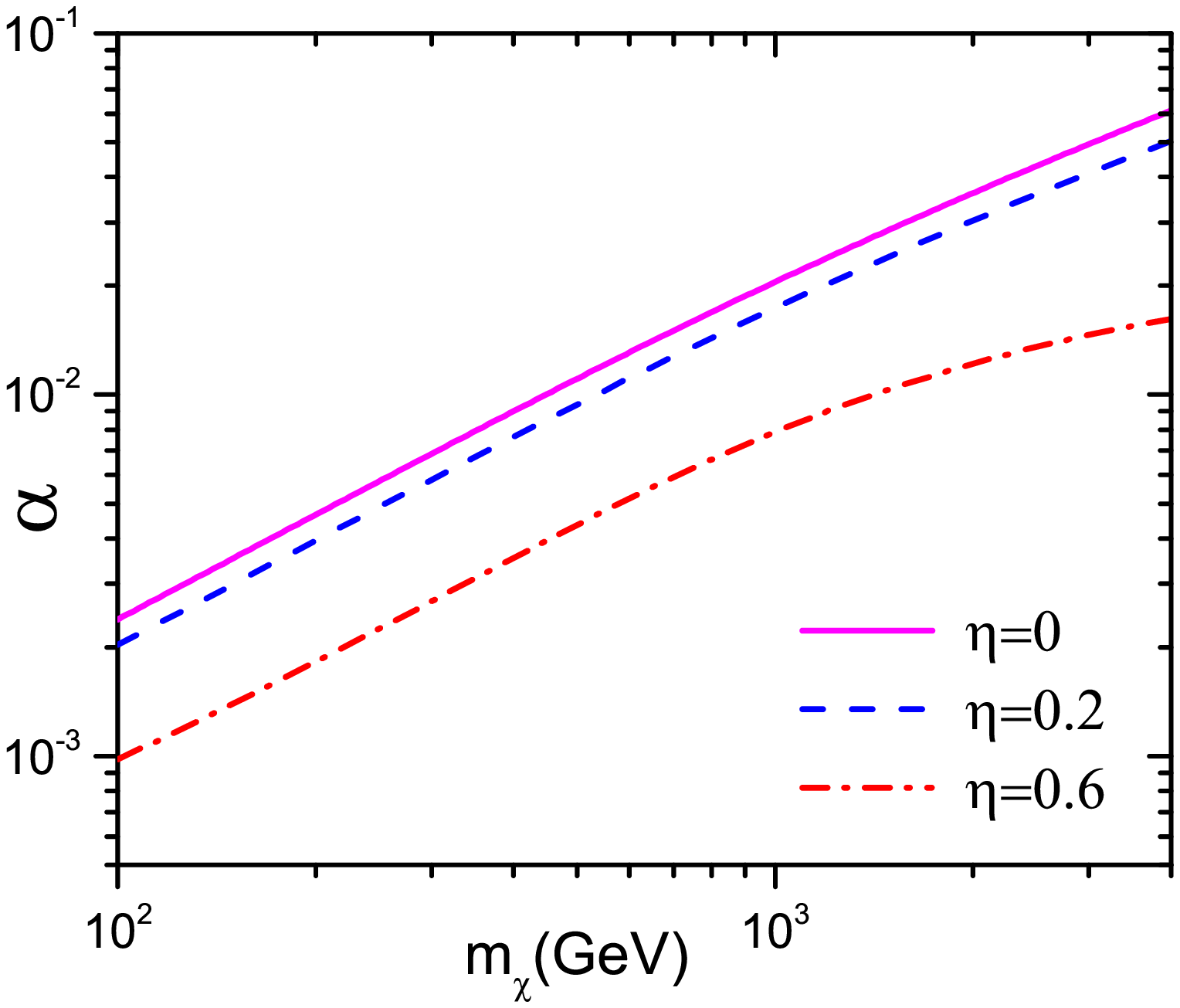}
\includegraphics[width=0.48\textwidth]{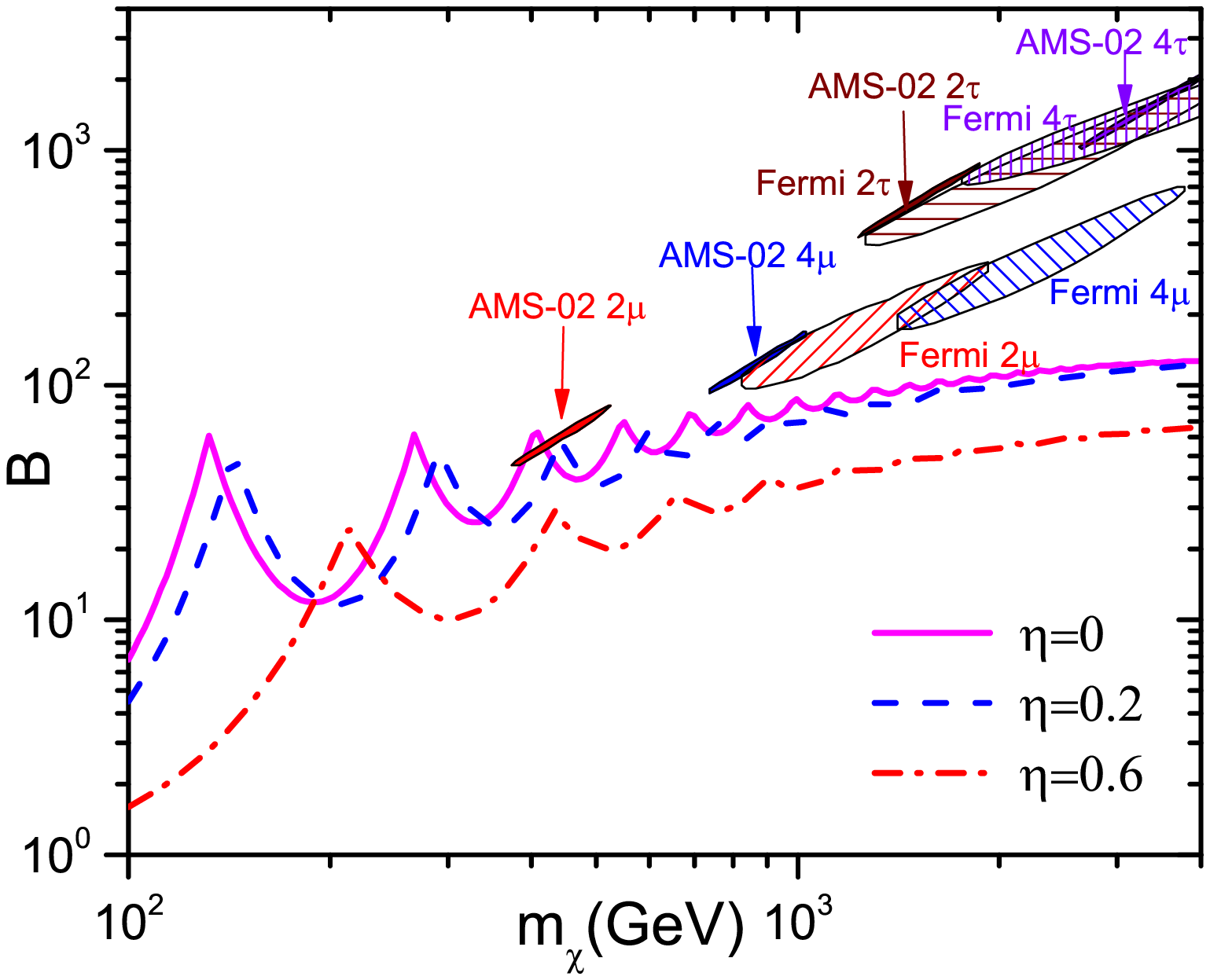}
\caption{
(Left)  Values of $\alpha$ constrained by the DM relic density
as a function of DM particle mass $m_{\chi}$
for  $\eta=0$, 0.2 and 0.6, respectively,
in the case where $\phi$ is a vector boson. 
(Right) Allowed  values of boost factor $B$
as a function of $m_{\chi}$. 
The favoured regions at $99\%$ C.L. from a global fit to the data of 
AMS-02 and Fermi-LAT are also shown~\cite{Jin:2013nta}. 
The mass of the vector force-carrier is fixed at $m_{\phi}=0.25$ GeV.
}\label{fig:vector}
\end{center}
\end{figure}

If the force carrier is a vector gauge boson,
the induced  potential from the multiple exchange of $\phi$ between the 
DM particles is of Yukawa type in \eq{eq:Yukawa}. 
For a vector force carrier, 
the DM particles can annihilate into $\phi\phi$ through 
$t$-channel diagram as shown in the right panel of  \fig{fig:diagrams}, 
which is an $s$-wave process.
The corresponding cross section reads
\begin{equation}\label{eq:vector}
\left( \sigma_{\phi\phi} v_{\text{rel}}\right)^{\text{vec}}_0= \frac{\pi \alpha^2}{m^2_{\chi}}  .
\end{equation}
According to \eq{eq:sigmaTotal}, 
for a given value of $\eta$, 
from calculating the DM thermal relic density and matching it to
the observed value in \eq{eq:omegahExp}, 
one can obtain the allowed values of the coupling $\alpha$
as a function of DM particle mass. 
The results are shown in the left panel of \fig{fig:vector}.
At $m_{\chi}\approx 460$ GeV, for $\eta=0$, the allowed coupling $\alpha$ is $~0.01$.
For larger $\eta$, the allowed $\alpha$ in general becomes smaller,
as the cross section $(\sigma_{\phi\phi}v_{\text{rel}})_{0}$ has to be smaller.
Making use of  the allowed values of $\alpha$, 
the allowed values of the boost factor $B$ are calculated,
and shown in the $(m_{\chi}, B)$ plane
in  the right panel of \fig{fig:vector}, 
together with the regions favoured by 
the AMS-02 and Fermi-LAT experiments at $99\%$ C.L.
from a global fit assuming a conventional astrophysical background~\cite{Jin:2013nta}.
The case where $\eta=0$ corresponds to 
the case previously discussed in Ref.~\cite{Feng:2010zp}, 
and our conclusion is in good agreement with theirs.  
As can be seen in the figure, 
in this case, 
the Sommerfeld enhancement can only 
marginally explain the data of AMS-02, which requires that
the enhancement should be  in the resonance region.
Since both the $\bar\chi\chi\to \phi\phi$ and $\bar\chi\chi\to X$ are 
$s$-wave processes, 
for nonvanishing $\eta$,  
even stronger upper bounds on the boost factor 
are obtained for larger values of $\eta$. 

\subsection{Scalar  force-carrier}
\begin{figure}[htb]
\begin{center} 
\includegraphics[width=0.48\textwidth]{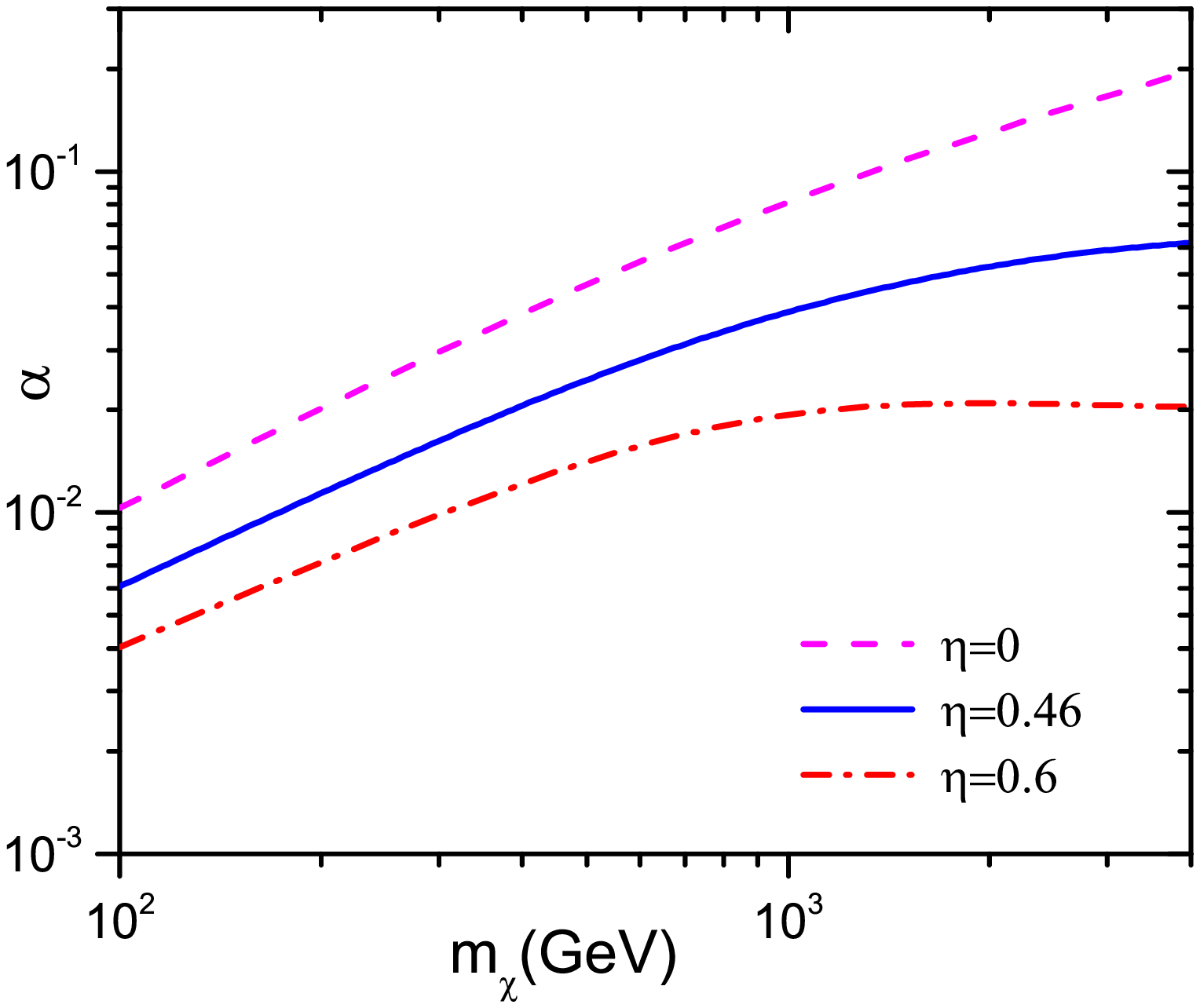} 
\includegraphics[width=0.48\textwidth]{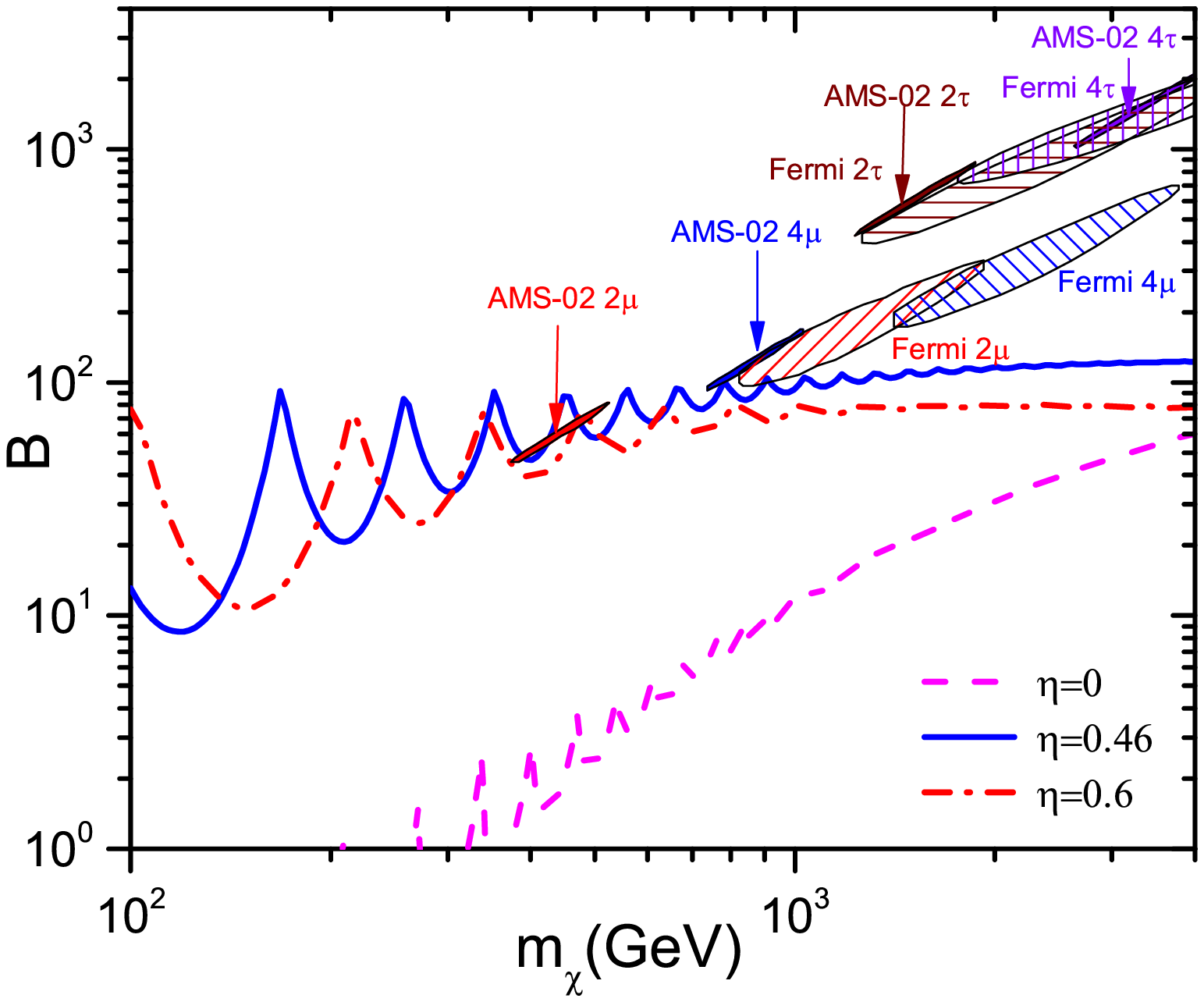} 
\caption{ 
The same as \fig{fig:vector}, but for the case where the force-carrier $\phi$
is a scalar particle.
}\label{fig:scalar}
\end{center}
\end{figure}

\begin{figure}
\begin{center}
\includegraphics[width=0.48\textwidth]{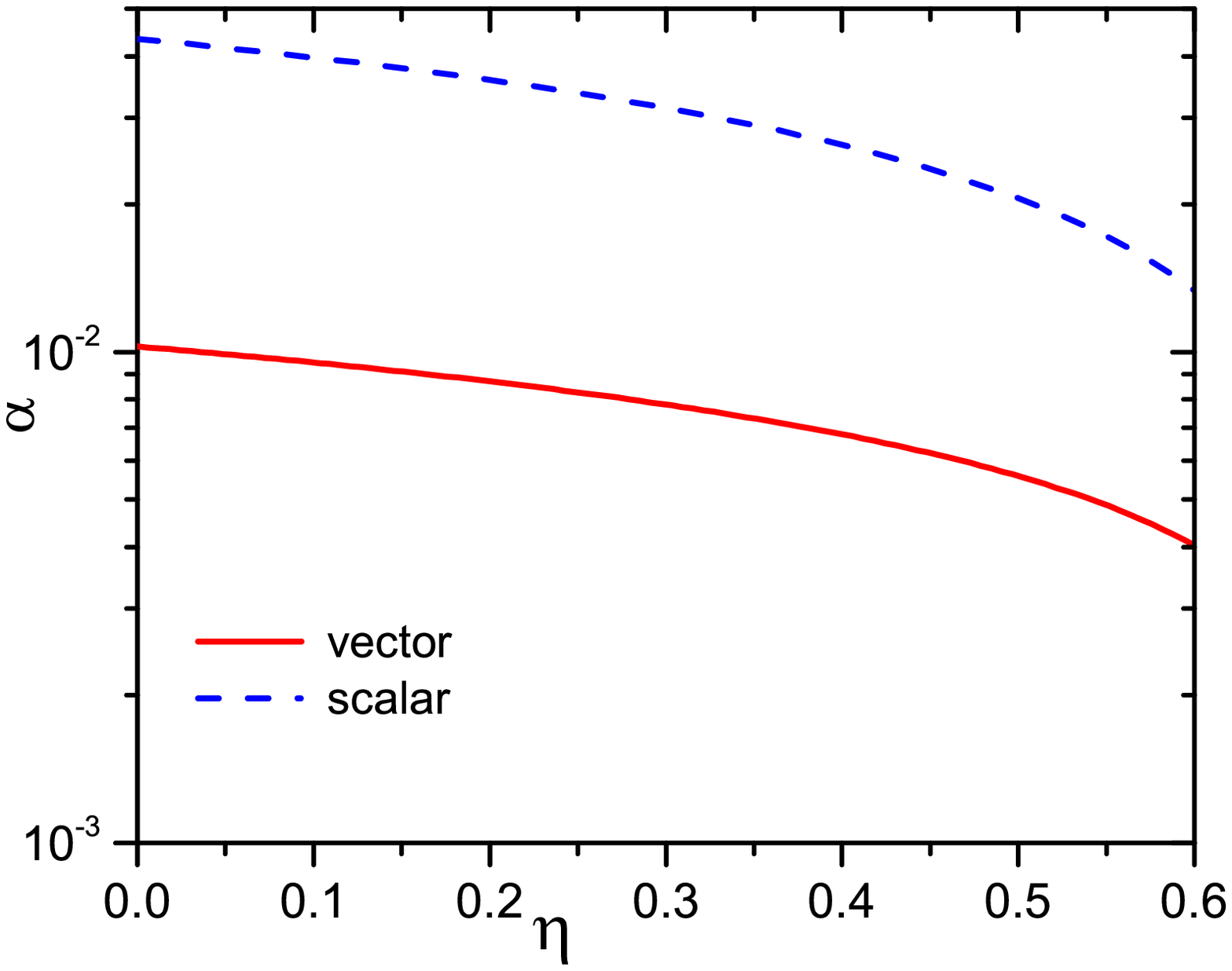}
\includegraphics[width=0.48\textwidth]{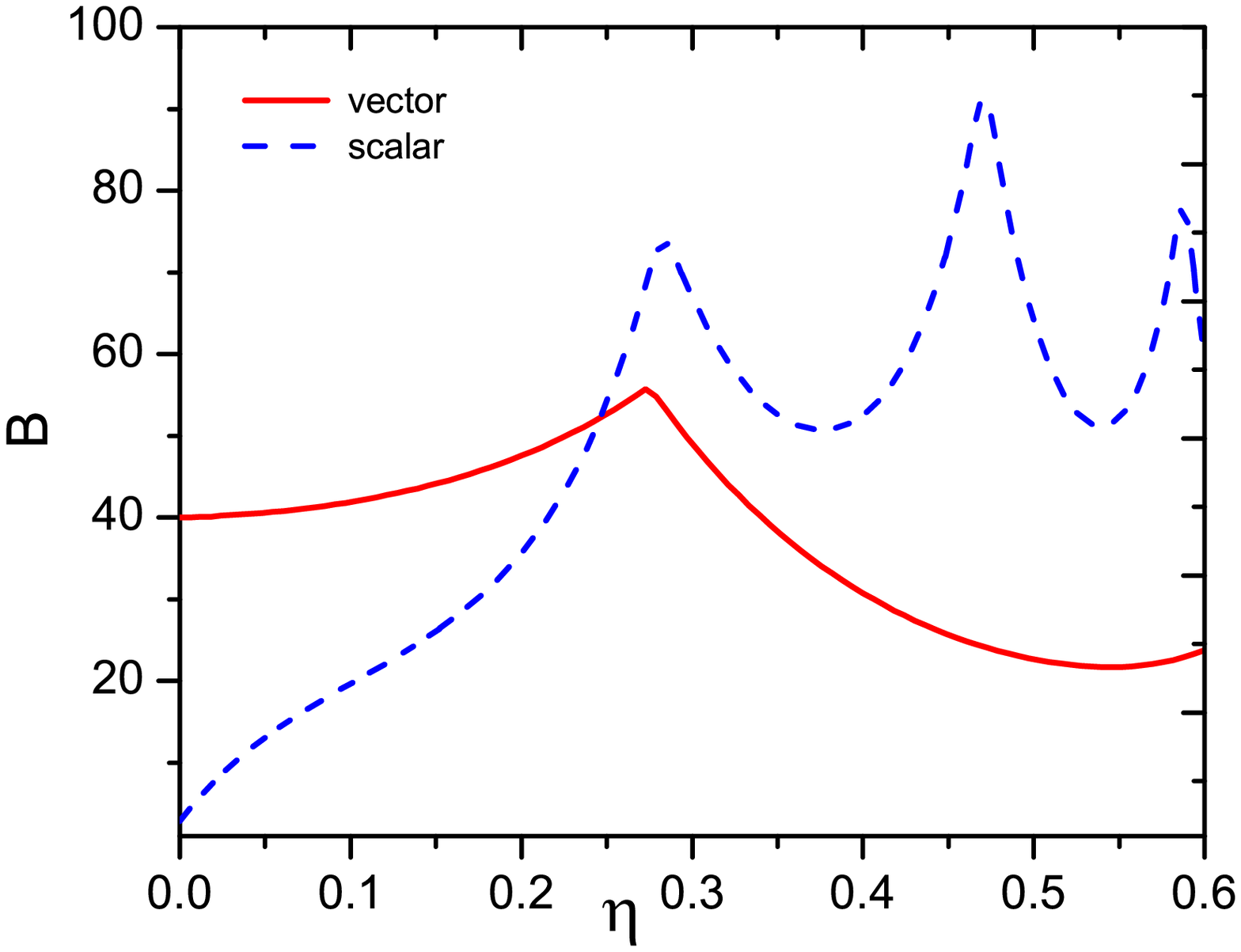}
\caption{ Comparison between the cases with vector and  scalar
force-carriers  on the allowed values of $\alpha$ (left)
and boost factor $B$ (right) as a function of $\eta$.
The masses of the DM particle and the force-carrier are
fixed at $m_{\chi}=460$ GeV and $m_{\phi}=0.25\mbox{GeV}$,
respectively.
}\label{fig:bf_eta}
\end{center}
\end{figure}

In the case where 
the force carrier is a scalar particle, 
the $t$-channel annihilation $\bar\chi\chi\to \phi\phi$ is a $p$-wave process which
has a velocity-dependent cross section
\begin{equation}\label{eq:scalar}
\left( \sigma_{\phi\phi} v_{\text{rel}}\right)^{\text{sca}}_0
= \frac{3\pi \alpha^2 }{8m^2_{\chi}} v^2_{\text{rel}}  .
\end{equation}
In principle,
there could  exist non-negligible cubic  self-interactions 
between the force-carriers of the form $ -\mu \phi^{3}/3!$, 
which leads to  an additional $s$-channel two-body annihilation.
The total annihilation cross section is modified as
$ (\sigma_{\phi\phi}v_{\text{rel}})^{\text{sca}}_{0}=(3\pi\alpha^{2}/(8m_{\chi}^{2}))(1-5\xi/18+\xi^{2}/48) v_{\text{rel}}^{2}$
with $\xi=\mu/(2m_{\chi }\sqrt{\alpha \pi})$~\cite{Chen:2013bi}.  
In this work, 
for simplicity, 
we only consider the case where  $\xi \ll 1$,
namely, the $t$-channel diagram dominates.
Compared with  \eq{eq:vector}, 
the cross section of DM annihilating into the scalar force-carriers
in \eq{eq:scalar}  is suppressed by 
both the prefactor $3/8$ and 
the small relative velocity $v^{2}_{\text{rel}}\sim 0.1$ at  freeze out. 
The corresponding constraint on  $\alpha$ from the DM thermal relic density 
is expected to  be weaker. 
In the left panel of \fig{fig:scalar}, 
we show the allowed value of $\alpha$ as a function of DM particle mass
from  the constraint of DM thermal relic density. 
In the numerical calculations, 
the  Sommerfeld enhancement for the  $p$-wave process in \eq{eq:scalar}  is also included. 
For $\eta=0$,  at $m_{\chi}=500$~GeV, the allowed value of coupling $\alpha$
is $\sim 0.02$
which is about a factor of two larger than that in the case with a vector force-carrier. 
In the right panel of \fig{fig:scalar},
the allowed boost factors are shown 
for three different choices of $\eta=0$, 0.46 and 0.6, respectively.
Contrary to the vector force-carrier case,  
for $\eta=0$, 
the allowed boost factor is very small,
which is due to the fact that in this case the  $p$-wave 
annihilation $\bar\chi\chi\to \phi\phi$ for scalar force-carriers
is velocity-suppressed after including the Sommerfeld enhancement.
For nonzero $\eta$, 
the allowed boost factor becomes larger.
However, the boost factor does not increase monotonically with increasing $\eta$.
We find that the maximally allowed boost factor at $m_{\chi} \approx 460$ GeV
corresponds to $\eta\approx 0.46$, 
which  can be  consistent with  that  favoured by the AMS-02 data for 
DM annihilating into $2\mu$ final states,
but is not large enough to account for other 
final states such as $2\tau$ and $4\mu$.
In \fig{fig:bf_eta},  we show how the values of $\alpha$ and the boost factor $B$ depend
on the value of $\eta$ for a fixed $m_{\chi} \approx 460$ GeV
for both vector and scalar force-carrier cases. 
As seen in the figure, for $\eta\approx 0.46$, 
the  Sommerfeld enhancement is close to a resonance, 
which leads to a relatively  large boost factor shown in \fig{fig:scalar}.
The effect of resonance is less significant for vector force-carrier case.

\subsection{Pseudoscalar force-carrier}

The interaction between 
a pseudoscalar force-carrier particle $\phi$ and a fermionic   DM particle $\chi$
is of  the form $i\bar{\chi}\gamma^{5}\chi\phi$,
which results in a  spin-dependent potential 
\begin{equation}\label{eq:tensor_pot}
V\left(r\right)=-\frac{\alpha}{4m_{\chi}^{2}r^{3}}e^{-m_{\phi}r}
\left[
3\left(\hat{\mathbf{r}}\cdot\hat{\mathbf{s}}_{1}\right)\left(\hat{\mathbf{r}}\cdot\hat{\mathbf{s}}_{2}\right)-\hat{\mathbf{s}}_{1}\cdot\hat{\mathbf{s}}_{2}
\right]  ,
\end{equation}
where $\hat{\mathbf{r}}$ is 
the unit  vector along the direction of the relative distance 
$\mathbf{r}$ of  the two annihilating DM particles,  and 
$\hat{\mathbf{s}}_{1,2}$  are the unit  vectors of  the spin orientation of 
the  DM particles.
This type of potential is known as the tensor-force potential in nuclear physics and 
is similar to  the potential induced by interaction of two electric dipoles. 
The Sommerfeld enhancement from this type of potential was discussed previously in Ref.\cite{Bedaque:2009ri}
without considering the constraints from DM thermal relic density.

The spin-dependent part of the potential can be rewritten as: 
$3\left(\hat{\mathbf{r}}\cdot\hat{\mathbf{s}}_{1}\right)\left(\hat{\mathbf{r}}\cdot\hat{\mathbf{s}}_{2}\right)
-\hat{\mathbf{s}}_{1}\cdot\hat{\mathbf{s}}_{2}
=\pm(3\cos^{2}\theta-1)$ for $\hat{\mathbf{s}}_{1}=\pm\hat{\mathbf{s}}_{2}$,
where $\theta$ is the angle between 
vectors $\hat{\mathbf{r}}$ and  $\hat{\mathbf{s}}_{1}$. 
The induced long-range force can be either attractive or repulsive, 
depending on the relative direction and the spin orientation of the annihilating DM particles.
In the case where the force is attractive (repulsive),  
Sommerfeld enhancements (suppressions) of DM annihilation can occur. 
Although for  unpolarized inital states of DM particles,
the  chances are equal  for the forces to be attractive or repulsive,
the net effect of the Sommerfeld enhancements and suppressions on the 
DM annihilation rate can be nonzero,
as  the enhancement of the annihilation rates can be dominant.
In this work, we consider the case where 
the  spins of two annihilating DM particles are parallel, i.e.,
$\hat{\mathbf{s}}_{1}=\hat{\mathbf{s}}_{2}$,
and calculate  the corresponding   Sommerfeld enhancement factors. 
The Sommerfeld suppression in the antiparallel  case is taking into account by 
adding an overall suppression factor $1/2$ to the boost factor $B$, 
which corresponds to the maximal suppression effect.

The potential matrix in the angular moment space  can be 
written in terms of the Wigner $3-j$ symbol as 
\begin{equation}\label{eq:dipole}
V_{\ell\ell'}\left(r\right)=
-\frac{\alpha e^{-m_{\phi}r }}{m_{\chi}^{2}r^3}
\sqrt{(2\ell+1)(2\ell'+1)}
\left(\begin{array}{ccc}
\ell & 2 & \ell' \\
0 & 0 & 0
\end{array}\right)^{2} .
\end{equation}
The off-diagonal elements of $V_{\ell\ell'}$ are nonvanishing for any
$\ell$ and $\ell'$ satisfying $|\ell'-\ell|=2$.
Thus the Schr$\ddot{\mbox{o}}$dinger equations for different partial waves are all  coupled together.
We solve the coupled Schr$\ddot{\mbox{o}}$dinger equations using
the Born-Oppenheimer adiabatic approximation. 
In this approach,  
a spatial-dependent rotation matrix $U_{i \ell}(r)$ is introduced to   
locally diagonalize the sum of the potential and the centrifugal term
in the  Schr$\ddot{\mbox{o}}$dinger equation at the position $r$ 
\begin{eqnarray} \label{eq:rotate}
H\left(r\right)_{ij}=U_{i\ell}(r) 
\left[ 
m_{\chi} V_{\ell\ell'}(r)+\frac{\ell(\ell+1)}{r^{2}}\delta_{\ell\ell'}
\right]
U^{T}_{\ell' j}(r)  ,
\end{eqnarray}
where $H(r)$ is a diagonal matrix. 
In this basis, 
the  only off-diagonal terms  in the Schr$\ddot{\mbox{o}}$dinger equation are proportional to $dU/dr$ or $d^2 U/dr^2$.  
For slowly varying potential  ( in comparison with the Compton
wave length of the particle falling into the center ) the terms proportional to 
 $dU/dr$ and $d^2 U/dr^2$  are relatively small 
and  can be neglected as a first order approximation.
Under this adiabatic approximation, 
the Schr$\ddot{\mbox{o}}$dinger equations 
for the rotated wave function $\phi_i\left( r \right)= U_{i\ell}(r) \chi_{\ell} \left(r \right) $
are decoupled, and have the simple form~\cite{2009JPhB...42d4017R}
\begin{equation} \label{eq:dipole_shro}
\frac{d^{2}}{dr^{2}}\phi_{i}
- H(r)\phi_{i} +k^{2} \phi_{i} \simeq 0  ,
\end{equation}
which  can be solved easily with the rotated  boundary conditions.
The solutions and the Sommerfeld
enhancement factors for each partial wave are obtained by performing 
an inverse rotation back into the original basis with definite angular momentum. 
Note that in the limit $ r \rightarrow \infty $ 
the centrifugal term $\ell(\ell+1)/r^2$ dominates over $V_{\ell\ell'}$.
The rotation matrix in this limit is a unit matrix. 
In numerical calculations, we consider the angular mumemtum up to $\ell=8$, thus
$V_{\ell\ell'}$ is a matrix of dimension-nine. We find good stability in the solutions of the
wave functions with lowest indices $\phi_{1,2}$.

It is known in quantum mechanics that 
for an attractive potential scaling with distance  
as $r^{-s}$ with $s \geq 2$,
the  wave function is not well-defined (divergent) at the origin. 
A procedure of regularization of this type of potential has to be introduced,
which represents the nonfactorizable contributions from the short-distance
(for a review, see Ref \cite{Frank:1971xx} ).  
In this work, we adopt a commonly used regularization scheme 
\begin{equation}\label{eq:cut-off}
V(r)\to V(r+r_{0})  ,
\end{equation}
where $r_{0}$ is a cut-off parameter. 
In the generic case, 
the pseudoscalar induced potential can be regularized as 
$r^{-3} \to r^{-\beta}(r+r_{0})^{\beta-3}$ with $\beta < 2$. 
The regularization scheme in \eq{eq:cut-off} corresponds to the case where $\beta=0$.
We have also performed  calculations for the case of $\beta=1$ and 
find no significant changes in the conclusions. 

\begin{figure}[htb]\begin{center}
\includegraphics[width=0.49\textwidth,]{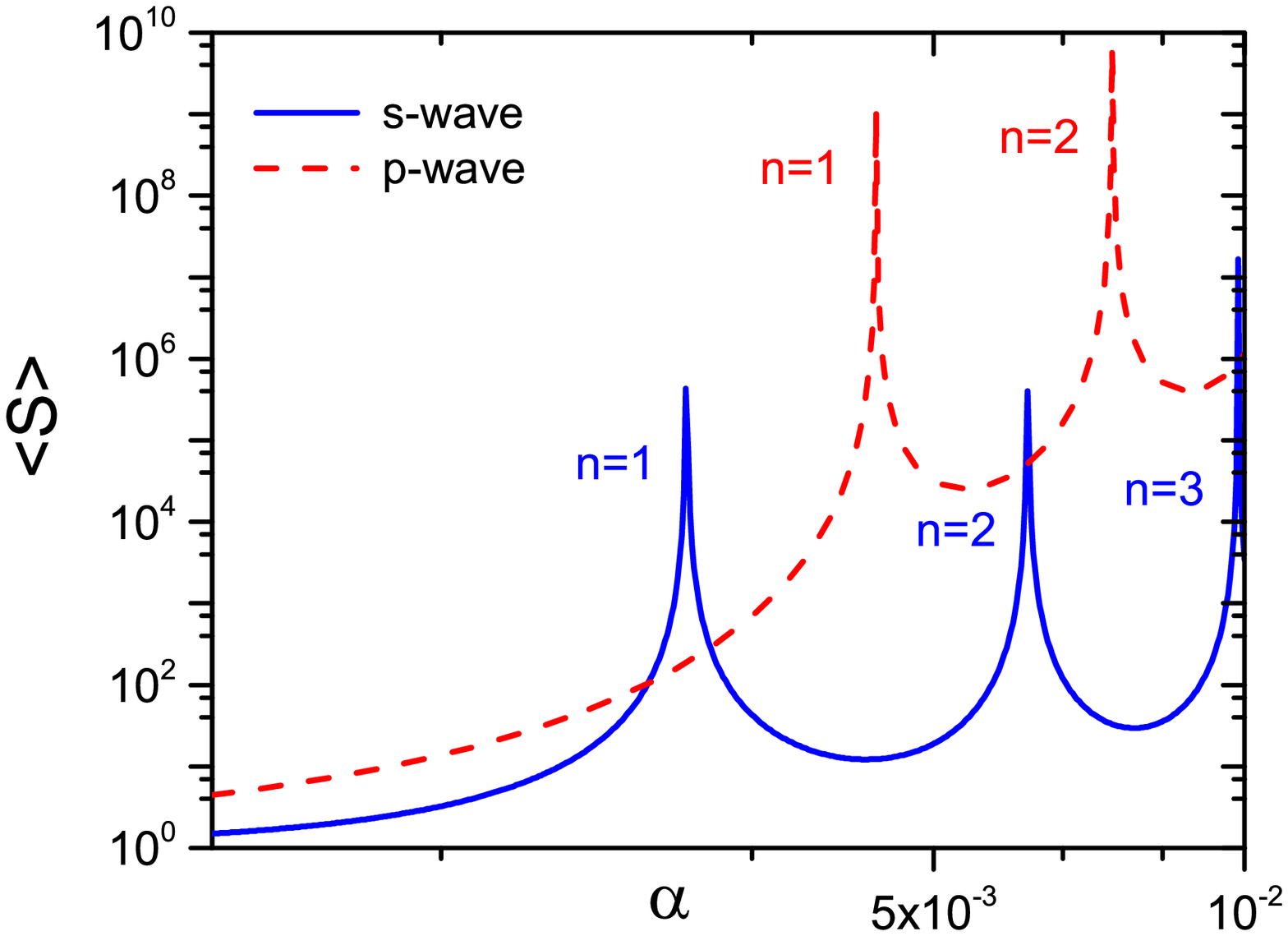}\includegraphics[width=0.49\textwidth,]{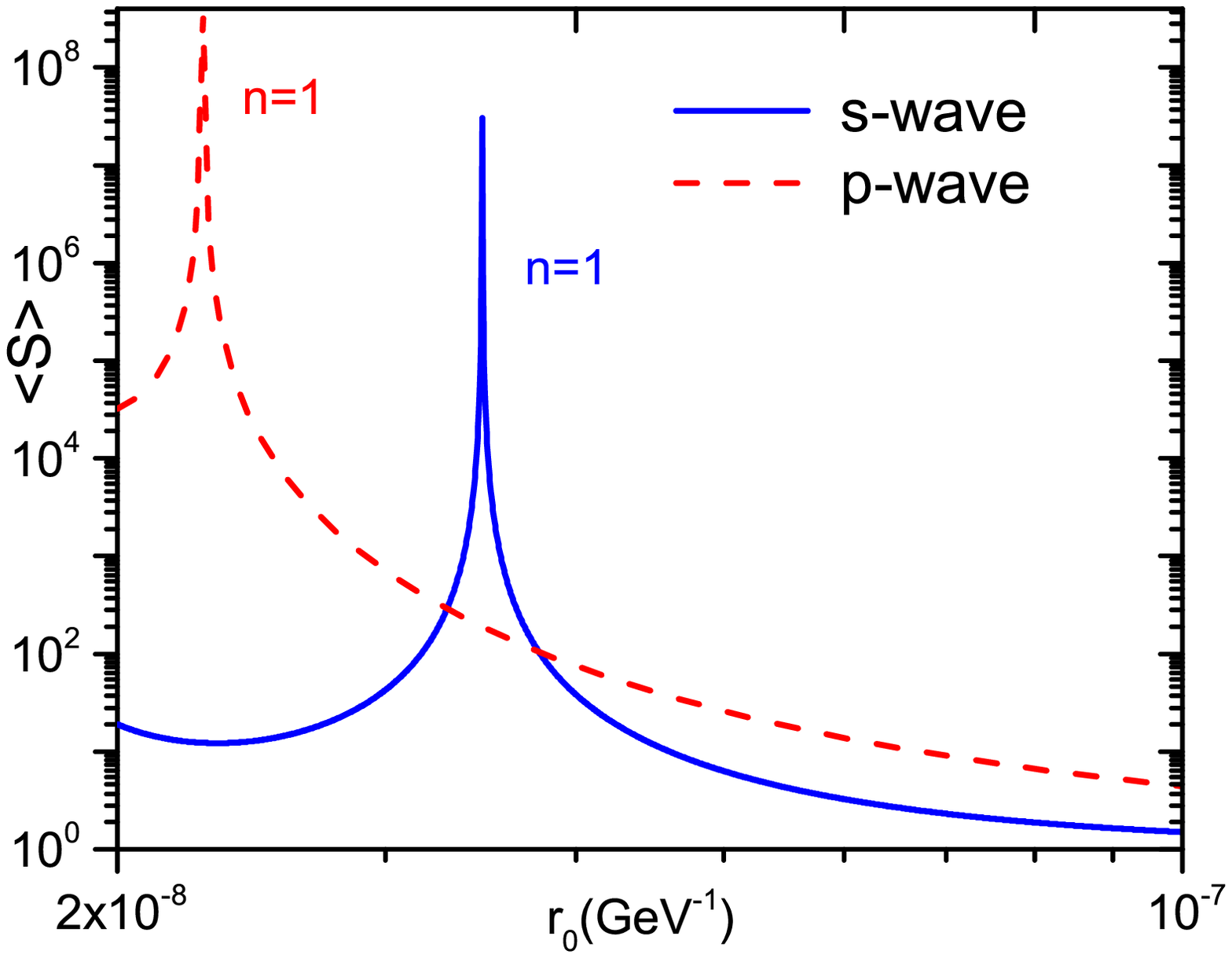}
\caption{ $s$-wave and $p$-wave thermally averaged Sommerfeld 
enhancement factors $\langle S \rangle$ as a function of $\alpha$ (left) and the cut-off scale
$r_{0}$ (right) for $x=x_{\text{now}}$. 
In the left panel, $r_{0}$ is fixed at $1.0\times 10^{-7} \text{ GeV}^{-1}$,
while in the right panel $\alpha$ is fixed at 0.001. 
The masses of the DM particle and the force-carrier  are fixed at $m_{\chi}=1000\mbox{ GeV}$ and  
$m_{\phi}=0.25\mbox{ GeV}$, respectively.
 }\label{fig:salp_cut}
\end{center}\end{figure}

\begin{figure}
\begin{center}
\includegraphics[width=0.60\textwidth]{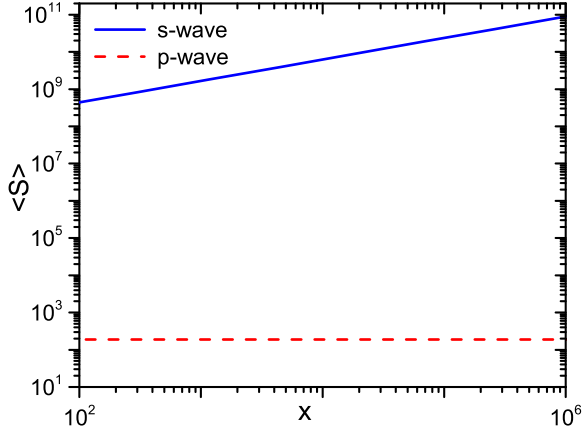}
\caption{Temperature  dependence o f
the thermally averaged Sommerfeld
enhancement factor in the $s$-wave resonance region,
corresponding to the resonance point with $r_0=1.0\times 10^{-7} \text{ GeV}^{-1}$, 
$\alpha=2.880\times 10^{-3}$  and $n=1$ as shown in \fig{fig:salp_cut}.
Other parameters are $m_{\chi}=1000\mbox{ GeV}$, and $m_{\phi}=0.25\mbox{ GeV}$.
The $p$-wave Sommerfeld enhancement factor is also shown, which show no
visible temperature dependence as it is off-resonance for the chosen parameters.
}\label{fig:sxf}
 \end{center}
\end{figure}

In \fig{fig:salp_cut}, 
we show how the thermally averaged Sommerfeld enhancement factor 
$\langle S\rangle$ depends on the coupling strength $\alpha$ and 
cut-off parameter $r_{0}$ for 
both $s$- and $p$-wave annihilation at the temperature 
$x=x_{\text{now}}$.
Similar to the case with Yukawa potential, 
in some regions of parameter space, 
resonant Sommerfeld enhancement occurs, 
which corresponds to the formation of zero-energy bound states.
We find that at the resonance points, the parameters $\alpha$, $r_{0}$,
and $m_{\chi}$ satisfy the following approximate relation
\begin{equation}\label{eq:resonanceCondition}
\alpha\approx \frac{r_{0}}{R} n m_{\chi},
\end{equation}
where $n=1,2,3,\dots$, and  $R\approx 0.0347 \ (0.0227)$ for $s(p)$-wave
annihilation. 
As expected, 
the enhancement factors become larger 
with increasing $\alpha$ and decreasing $r_{0}$. 
As shown in \eq{eq:boostFac}, 
the boost factor depends only on 
the size of the Sommerfeld enhancement at the present day relative  to that 
at the time of freeze out. 
For the regularized singular potentials,  
only at the resonance points, 
the Sommerfeld enhancement factor depends  significantly 
on the temperature.
In \fig{fig:sxf}, we show the thermally averaged Sommerfeld enhancement factor
of pseudoscalar induced potential for $s$- and $p$-wave cases. 
The parameters are chosen to be 
$\alpha=2.88\times 10^{-3}$, $r_{0}=1.0\times 10^{-7}\text{ GeV}^{-1}$,
and $m_{\chi}=1$ TeV, 
which corresponds to the  $s$-wave resonance point with $n=1$. 
From the figure, 
one sees that the thermally averaged Sommerfeld enhancement factor
at $x\approx x_{\text{now}}$ can be a few hundred times larger than that at 
freeze out $x\approx x_{f}$. 
For the same parameter set, 
the $p$-wave annihilation is not at the resonance point, 
thus there is no relative enhancement towards low temperatures.
This feature  is  similar to the case with a  spherical well potential
$V_{\text{well}}\left( r \right)=-V_0\theta\left(r-r_0\right)$ 
with $\theta(x)$  the Heaviside step function.
For the spherical well potential, 
the Sommerfeld enhancement factor for $s$-wave annihilation 
is given by \cite{Hannestad:2010zt} 
\begin{equation}\label{eq:smf_well_sw}
S^{\text{well}}_{0}(v_{\text{rel}})=\frac{1}{1-\frac{V_0}{V_0+m_{\chi}v^2_{\text{rel}}/4}
\sin^2(r_0\sqrt{4 m_{\chi}V_0 + m^{2}_{\chi}v^2_{\text{rel}}})}  .
\end{equation}
For a deep well $V_0 \gg m_{\chi}v^2_{\text{rel}}/4$, 
in the resonant region, i.e.,
$r_{0} \sqrt{4 m_{\chi} V_{0}}\approx n+\pi/2$, 
one obtains  $S^{\text{well}}_{0}(v_{\text{rel}})\sim 4V_0/\left(m_{\chi}v^2_{\text{rel}}\right)$.
But when it is off-resonance, 
$S^{\text{well}}_{0}(v_{\text{rel}})\approx 1$.

\begin{figure}  
\begin{center}
\includegraphics[width=0.49\textwidth]{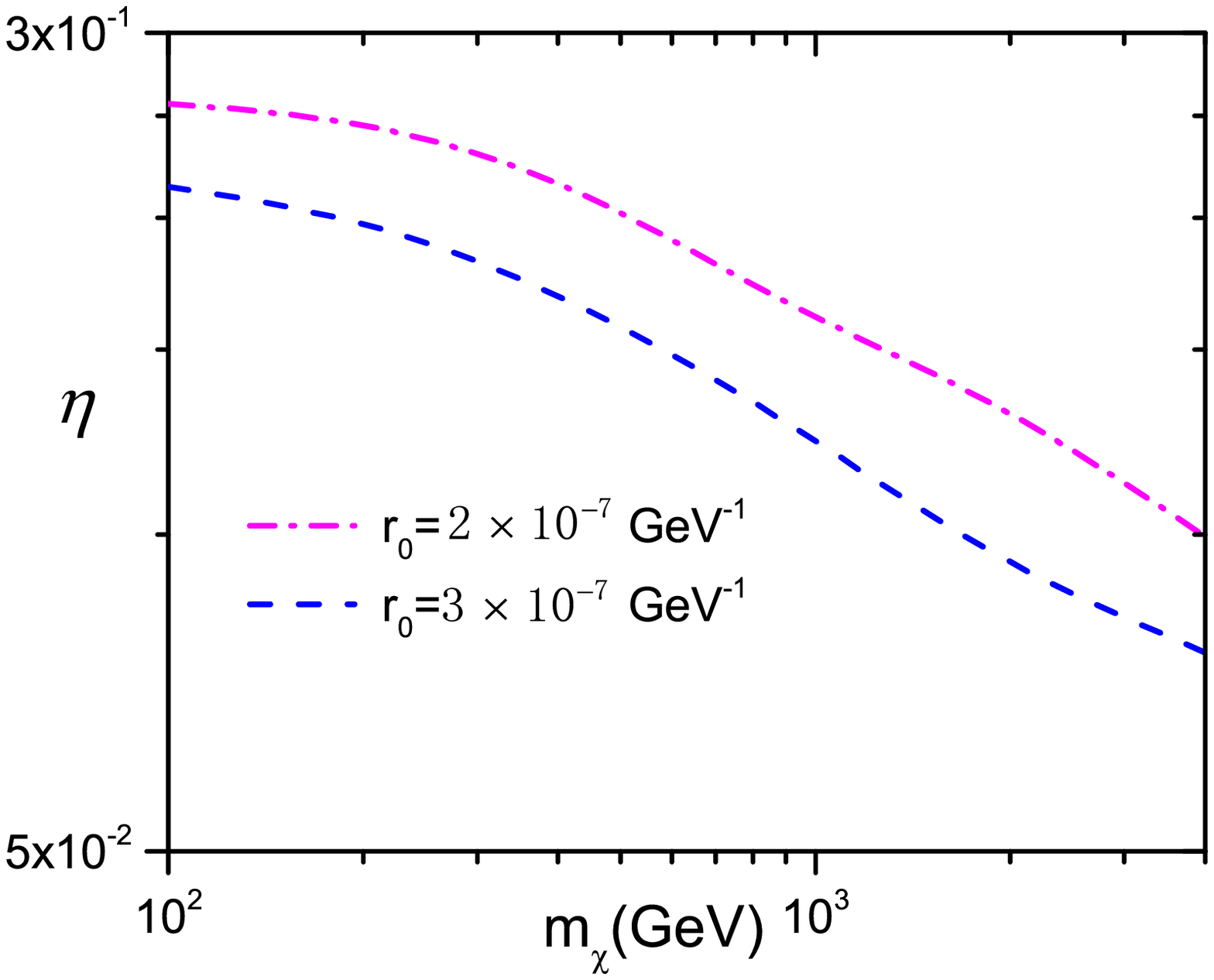}
\includegraphics[width=0.49\textwidth]{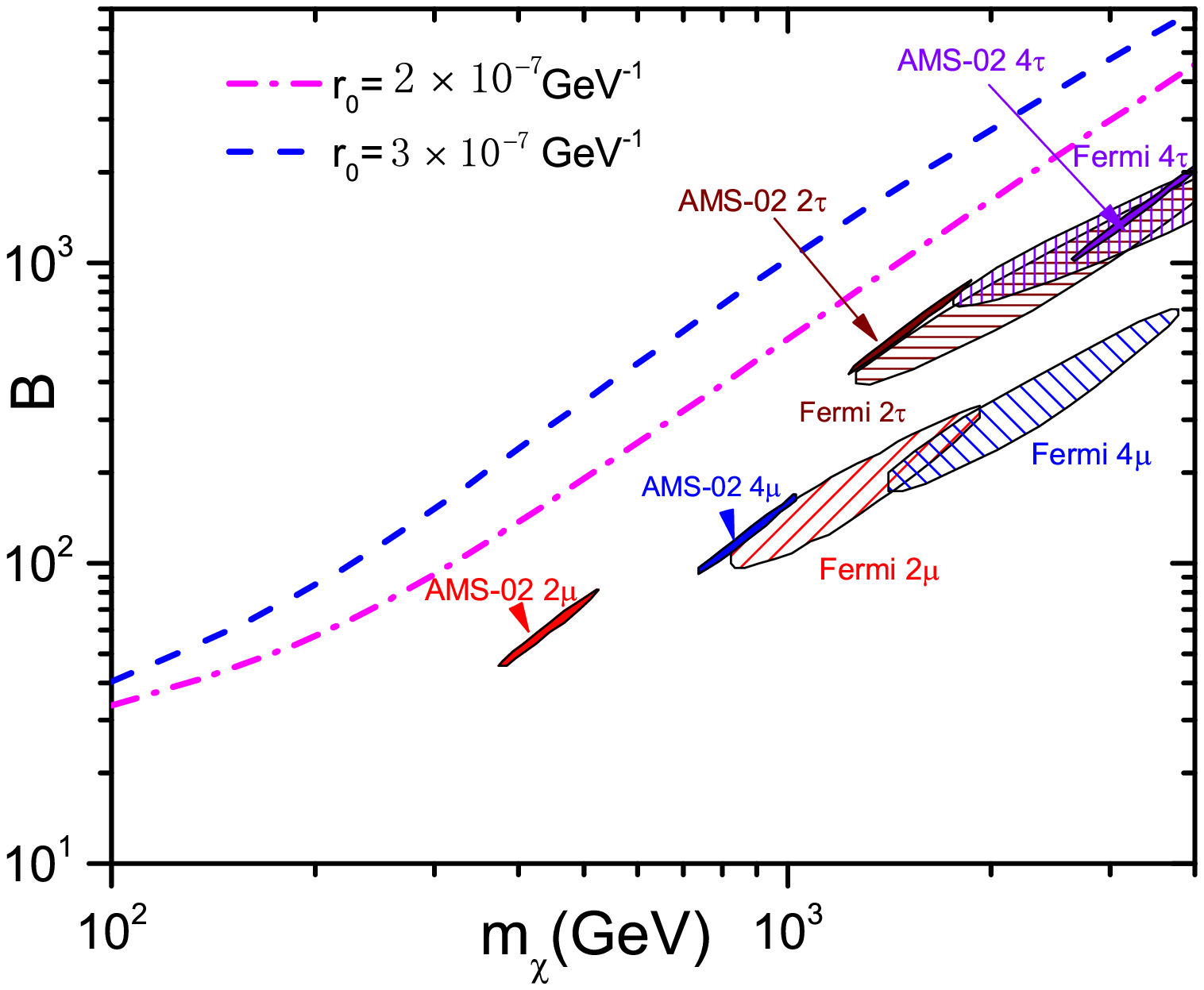}
\caption{(Left) Values of $\eta$ constrained by the 
DM thermal  relic density as a function of DM particle mass
with two different cut-offs $r_{0}=2.0 \times 10^{-7} \text{ GeV}^{-1}$ and
$3.0\times 10^{-7} \text{ GeV}^{-1}$.
(Right) The allowed boost factors as a function of DM particle mass.
The mass of pseudoscalar force-carrier  is fixed at 
$m_{\phi}=0.25\mbox{ GeV}$.
}\label{fig:pscalar}
\end{center}
\end{figure}

We proceed to discuss  the constraints on the Sommerfeld enhancement 
from DM thermal relic density.  
For the pseudoscalar force carrier $\phi$, 
the cross section for the  $t$-channel DM annihilation process $\bar{\chi}\chi \rightarrow 2\phi$ 
is given by
\begin{align}\label{eq:pscalar}
 \left( \sigma_{2\phi} v_{\text{rel}} \right)^{\text{ps}}_0  =\frac{\pi\alpha^{2}}{24m_{\chi}^{2}} v_{\text{rel}}^{2} .
\end{align}
Compared with the cross section of the scalar force-carrier case in \eq{eq:scalar}, 
it is smaller by a factor of  nine,
which results in a weaker constraint from thermal relic density. 
Since at the resonance points 
the coupling strength $\alpha$ is related to other parameters such as 
$r_{0}$ and $m_{\chi}$ through \eq{eq:resonanceCondition},
we show instead in the left panel of \fig{fig:pscalar} 
the constraints on the parameter $\eta$,
for two different cut-offs $r_{0}=2.0 \times 10^{-7} \text{ GeV}^{-1}$ and
$3.0\times 10^{-7} \text{ GeV}^{-1}$, respectively.
The decrease of  $\eta$ at larger  $m_{\chi}$,
is due to the increase of $\langle S\rangle (x_{f})$,
as $\alpha$ is related to $m_{\chi}$ through \eq{eq:resonanceCondition}.
For larger $r_{0}$, the value of $\alpha$ is larger, 
thus the required $\eta$ becomes smaller. 
The allowed boost factors  at the present day  are
shown in the right panel of \fig{fig:pscalar}. 
One sees that in the case with pseudoscalar force carrier,  
the allowed Sommerfeld enhancement factors can be 
large enough to account for  the excesses reported by
 AMS-02  and Fermi-LAT for a variety of final states such as $2\mu$, $2\tau$,
 $4\mu$ and $4\tau$, etc..

\section{Conclusions}\label{sec:conclusion}

Although 
the Sommerfeld enhancement has been considered as 
a mechanism for naturally increasing the DM annihilation cross section 
at low relative velocities, 
which is crucial to explain the current data of PAMELA, Fermi-LAT and AMS-02,
stringent  constraint can arise from the  DM thermal relic density, 
partially due to the annihilation of DM particles into the force-carriers 
introduced by this mechanism. 
We have shown that 
the effect of the Sommerfeld enhancement and 
the constraint from thermal relic density depend on
the nature of the force-carrier particle. 
For the force-carrier being a vector boson, 
the induced long-range potential is of Yukawa type,
and the process of $\bar\chi\chi\to \phi\phi$ is an $s$-wave process.
If $\phi$ is a scalar, 
the same process becomes a  velocity-suppressed $p$-wave process, 
which resulting in a weaker constraint.
If $\phi$ is a pseudoscalar, 
the induced long-range potential is a tensor force,
and $\bar\chi\chi\to \phi\phi$ is again a $p$-wave process.
We  have explored  and compared  the  Sommerfeld enhancements 
with these three type of force-carriers
under the constraint from DM thermal relic density.
The results show  that 
for vector boson force-carrier 
the Sommerfeld enhancement  can only 
marginally account for the AMS-02 data, 
for scalar force-carrier 
the  allowed Sommerfeld enhancement factor can be larger roughly by a factor of two,
while in the case of pesudoscalar force carrier, 
much larger enhancement can be obtained in the resonance region. 
The Sommerfeld enhancement may still be a viable mechanism to account for 
the current cosmic-ray lepton anomalies.   

\section*{Acknowledgments}
This work is supported in part by
the National Basic Research Program of China (973 Program) under Grants No. 2010CB833000;
the National Nature Science Foundation of China (NSFC) under Grants No. 10975170,
No. 10821504   
and No. 10905084;
and the Project of Knowledge Innovation Program (PKIP) of the Chinese Academy of Science.

\bibliographystyle{JHEP}
\bibliography{reference,misc}
\end{document}